\documentclass[final,5p,times,twocolumn]{elsarticle}
\usepackage{graphicx}
\usepackage{amssymb}
\usepackage{amsmath}
\usepackage{bbm}
\usepackage{bm}
\usepackage{lscape}
\usepackage{color}
\journal{Physica D}

\begin{document}
\begin{frontmatter} 

\title{Relay synchronization and control of dynamics in multiplex networks with unidirectional interlayer coupling }

\author[first]{Aiwin T Vadakkan\corref{aff}}
\affiliation[first]{Indian Institute of Science Education and Research Tirupati, Tirupati-517 619, India}
\cortext[aff]{Currently at University of Naples Federico II, Naples, Italy}

\author[second]{G. Ambika\corref{cor1}}
\affiliation[second]{Indian Institute of Science Education and Research Thiruvananthapuram, Thiruvananthapuram-695 551, India}
\cortext[cor1]{Corresponding Author\\ Email address: g.ambika@iisertvm.ac.in}

\begin{abstract} 
Multiplex networks provide a proper framework for understanding the dynamics of complex systems with differing types of interactions. This study considers different dynamical states possible in a multiplex network of nonlinear oscillators, with a drive layer and two identical response layers where the interlayer interactions are unidirectional. We report how the directionality in coupling can lead to relay synchronization with amplification in the two response layers through feedback from the middle drive layer. The amplitude of synchronized oscillations of response layers can be controlled by tuning the strength of interlayer coupling.   Moreover, we find the synchronization patterns that emerge in the response layers depend on the nature of interlayer coupling, whether feedback or diffusive, and the time scale or parameter mismatches between drive and response layers. Thus, the study indicates the potential for controlling and optimizing the dynamics of response layers remotely by adjusting the strength of interlayer coupling or tuning the dynamic time scale of the drive layer.
\end{abstract}


\end{frontmatter}

\section{Introduction}

The emergent dynamics of several real-world complex systems, having many interacting subsystems, can be effectively modelled using the framework of complex networks. When the subsystems are not identical in the dynamics or nature of their interactions, we can model them using multilayer networks so that the different layers can handle the dissimilarities \cite{Kivela2014}.  Multiplex networks are a special case of multilayer networks, where all layers have the same number of nodes, and the $i^{th}$ node of each layer is connected only to the $i^{th}$ node of the other layers  \cite{WANG2024110300}. This framework can be beneficial for studying different types of emergent dynamics  \cite{Saxena2012,Pranesh2023,Koseska2013,Zakharova2014,Sethia2014,Zakharova2014, Banerjee2015}, as in nervous systems at various organizational levels \cite{Makovkin2020}, social interaction networks with varying interactions  \cite{Deville2016}, and power grid networks having different loads for connections \cite{Pagani2013} etc. Additionally, this can add an extra layer of controllability to the system such that we can achieve a desired state in one of the layers by tuning the parameters of the other  \cite{VADAKKAN2024129842}.

Among the possible dynamical states in complex systems, synchronization is the most relevant and well-studied collective phenomenon due to its theoretical, biological, and technological significance. Thus, in both single and multilayer networks, synchronization is achieved in multiple ways \cite{8527648, Belykh2019,article,Anwar2023} with a variety of synchronous behaviours reported, such as frequency, in-phase and anti-phase \cite{dynamics}, complete, cluster \cite{Jalan2016, cluster}, explosive \cite{PhysRevE.88.042808, BAYANI2023113243, Zhang2015,Verma2022,Kumar2020}, intra layer \cite{Gambuzza2015,Rakshit2020}, and inter layer \cite{inter, Leyva2017, Pitsik2018} synchronization. Moreover, recent research has gone beyond pairwise interactions, focusing on higher-order interactions in multilayer networks that can efficiently handle real-world scenarios, like social networks or neuronal networks, where interactions occur among groups of three or more systems simultaneously. In this context, a few studies have explored the impact of higher-order interactions on synchronization in multilayer networks \cite{MAJHI2025144,PhysRevResearch.6.033003}.

Apart from these,  a recently reported phenomenon is the concept of relay synchronization in a multilayer network where two distant networks, which are not directly connected, can synchronize with each other with the help of an intermediate network acting as the relay \cite{Kuptsov2015, Bergner2012, Fischer2006, Banerjee2012}. One specific application in this context is the human brain network, where the thalamus acts as a relay between distant cortical areas through the thalamocortical pathways \cite{GOSAK20221, JI20231, Guillery2002, Sherman2007, Mitchell2014, Vlasov2017}. In recent studies, researchers have looked at the effects of network topology  \cite{Drauschke2020}, edge weights \cite{Anwar2021}, intralayer coupling \cite{Leyva2018} and even repulsive coupling \cite{Wei2024} in achieving relay synchronization. Also, the relay synchronization of chimeras occurs, where synchronization of the coherent domains in the first layer occurs with their counterparts in the third layer \cite{Sawicki2018}. In a specific context, relay synchronization improves when the distant layers have positive coupling while the relay layer has repulsive coupling  \cite{Wei2024}.

Most of the reported studies on relay synchronization have focused on bidirectional coupling in intra-layer and inter-layer regimes. Hence, the effects of the directionality of interlayer links in relay synchronization are much less explored. There are, in fact, many complex real-world systems like the brain that function with its component networks having unidirectional connectivity. We note interesting patterns like nonstationary and imperfect chimera states occur in two-layer neuronal networks with unidirectional interlayer connections  \cite{Li2019}.

We also note that most of the real-world complex systems have constituents that often evolve under nonidentical time scales  \cite{npg-30-481-2023}. In such cases, the framework of multiplex networks would be an effective approach to model them, with the interacting units separated into multiple layers depending on their time scales. Some such real-world scenarios with multiple time scales are temporal neural dynamics  \cite{Samek2016, Boyden2005}, chemical reactions \cite{Das2013}, hormonal regulation \cite{Radovick1992},  and population dynamics \cite{Bena2007}. Recent studies indicate that the time scale mismatch between the layers can generate dynamical states such as amplitude death  \cite{Gupta2019}, cluster synchronization \cite{Yang2017}, and frequency synchronization \cite{Kachhara2021} and revival of synchronized oscillations \cite{VADAKKAN2024129842}.

In this study, we explore the role of unidirectional coupling in achieving relay synchronization in a three-layer multiplex network. In this framework, the middle layer (L2) acts as the drive with inter-layer unidirectional coupling with two other layers, L1 and L3. Using the Stuart Landau oscillator with periodic limit cycle dynamics at the nodes, we study the onset of relay synchronization in the system. When the interlayer coupling is of feedback type, we find that the drive layer L2 induces relay synchronization in L1 and L3 with amplification of oscillations and in phase synchronization with the drive layer. We show how the amplitude of oscillations in the response layers can be controlled by adjusting the strength of interlayer coupling or tuning the dynamical time scale mismatch between the drive and the responses. When the inter-layer coupling is of the diffusive type, drive layer L2 induces complete synchronization on all three layers in the network. The time scale mismatch between layers can also be tuned to control the nature of dynamics on the response layers, giving rise to interesting patterns like quasi-periodic states in response layers.

\section{Relay synchronization in three layer multiplex network with unidirectional interlayer coupling}

\begin{figure}[!ht]
\includegraphics[width=0.35\textwidth]{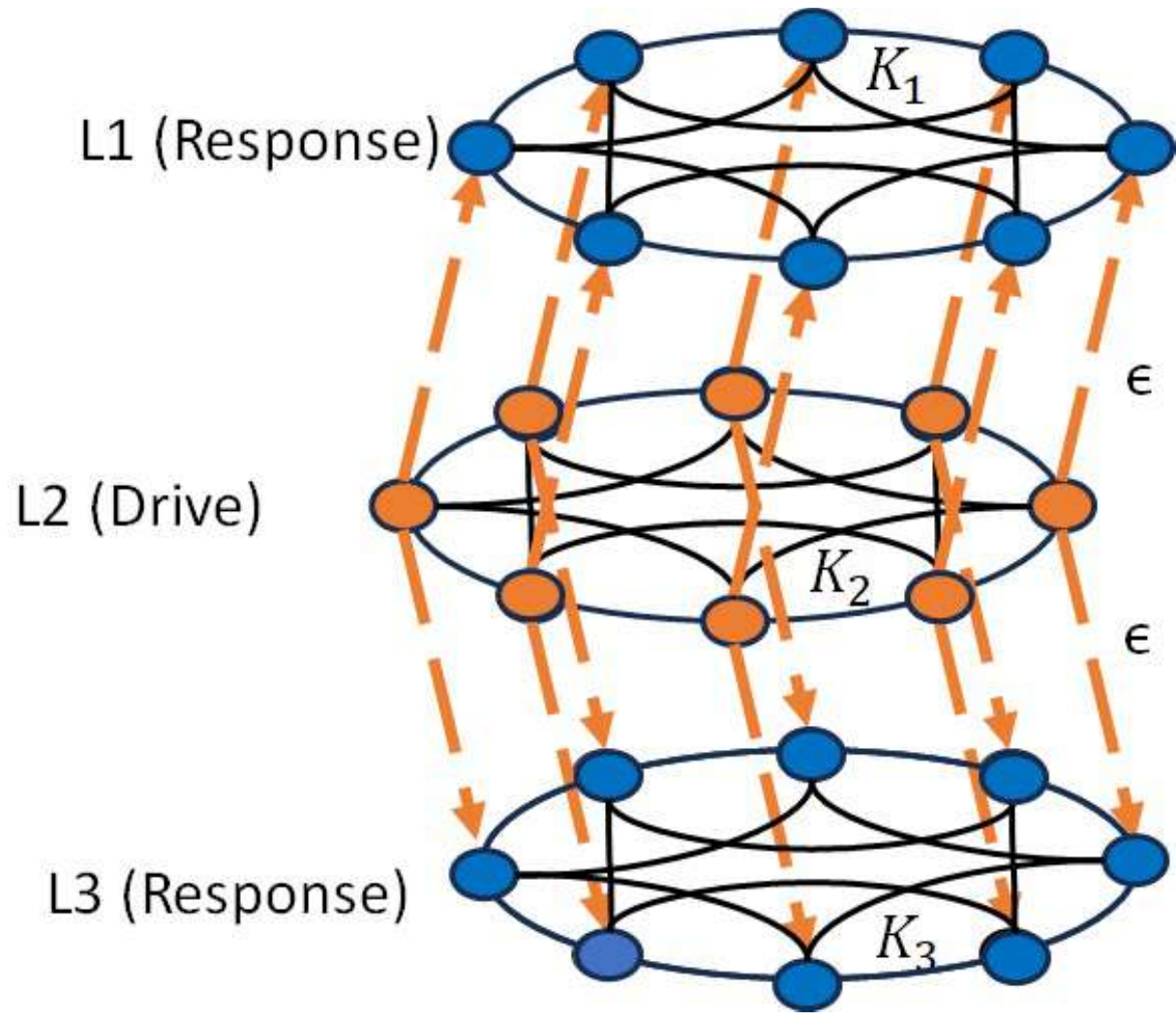}
\caption{Schematic of a three-layer multiplex network with unidirectional coupling from L2 (drive layer) to L1 and L3 (response layers). $\epsilon$ is the strength of unidirectional interlayer coupling between drive and response layers. K1, K2 and K3 are the strengths of intralayer couplings of L1, L2 and L3, respectively. }
\label{Fig1}
\end{figure}

We start with a three-layer multiplex network of Stuart- Landau (SL) oscillators with ring topology, where each layer has bidirectional intralayer diffusive coupling. The $N$ oscillators in the first (L1) and third layers (L3) are connected to the corresponding ones in the second layer (L2) via interlayer coupling of the feedback type to form a multiplex network. The interlayer coupling is unidirectional, indicating that L1 and L3 receive feedback from L2 through both x and y variables without any feedback input to L2, as shown in  Fig.~\ref{Fig1}. 

The dynamical equation for such a three-layer multiplex network is shown below.
 
\begin{align}
\dot{x}_{i1} &= (1-x_{i1}^2-y_{i1}^2)x_{i1} - \omega y_{i1} 
+ \frac{K_1}{2P_1}\sum_{j=i-P_1}^{i+P_1}(x_{j1}-x_{i1}) 
+ \epsilon x_{i2} \nonumber \\
\dot{y}_{i1} &= (1-x_{i1}^2-y_{i1}^2)y_{i1} + \omega x_{i1} +\epsilon y_{i2} \nonumber \\
\dot{x}_{i2} &= \tau\left[(1-x_{i2}^2-y_{i2}^2)x_{i2} - \omega y_{i2} 
+ \frac{K_2}{2P_2}\sum_{j=i-P_2}^{i+P_2}(x_{j2}-x_{i2})\right] \nonumber \\
\dot{y}_{i2} &= \tau\left[(1-x_{i2}^2-y_{i2}^2)y_{i2} + \omega x_{i2} \right]\nonumber \\
\dot{x}_{i3} &= (1-x_{i3}^2-y_{i3}^2)x_{i3} - \omega y_{i3} 
+ \frac{K_3}{2P_3}\sum_{j=i-P_3}^{i+P_3}(x_{j3}-x_{i3}) 
+ \epsilon x_{i2} \nonumber \\
\dot{y}_{i3} &= (1-x_{i3}^2-y_{i3}^2)y_{i3} + \omega x_{i3}+ \epsilon y_{i2} \nonumber \\
\label{eq1}
\end{align}

where the variables $x_i$ and $y_i$ for i = 1, 2,..., N define the dynamics of SL oscillators in all three layers, and $\omega$ is the inherent frequency of their limit cycle oscillations. The intralayer coupling strength $K_1$ ($K_3$) and the coupling range $P_1$ ($P_3$) control the interactions among the SL oscillators in the response layers L1 (L3), whereas $K_2$ and $P_2$ control the interactions in the drive layer (L2). In this scenario, P represents the range of coupling in each direction, with $P \in \{1, \frac{N}{2}\}$, where $P=1$ for local connections and $P=\frac{N}{2}$ for a global coupling, and $P$ is $1 < P < \frac{N}{2}$ for nonlocal coupling.

The unidirectional feedback coupling from drive to response governs the inter-layer interactions, and without loss of generality, we take their strengths to be equal to $\epsilon$. We also introduce a parameter $\tau$ that can decide the dynamical time scale mismatch between the drive and response layers, such that the response layers L1 and L3 can be made to evolve on a slower time scale than the drive layer L2 by choosing a value of $\tau > 1$.

Initially, we keep $\tau = 1$ so that the three layers evolve under identical time scales. The intrinsic dynamics has $\omega = 2$ for each layer with $N = 100$ systems. We consider a typical case of nonlocal coupling in all three layers with $P1 = P2 = P3 = 25$. We set the drive network with intra-layer coupling strength  $K2 = 5$ so that it has complete synchronization. First, we try the case with  $K1 = K3 = 0$ so that the response layers L1 and L3 do not have intra-layer synchronization when there is no drive from L2 for $\epsilon = 0$. 
The system of equations in Eq. (\ref{eq1}) are integrated for random initial conditions between (-1, 1) for $10^5$ iterations and after removing transients, the integrated outputs are used to check for synchronization and nature of emergent dynamics on the three layers of networks.

As we increase the inter-layer coupling strength ($\epsilon$) between the drive and responses, the response layers achieve intra- and inter-layer synchronization, with amplification relative to the drive. To detect the onset of intra-layer synchronization, we define the corresponding synchronization error over all the nodes in each layer as  

\begin{equation}
    S_{intra} =  \Biggl \langle \sqrt{\frac{1}{N}\sum_{i=1}^{N}[(x_{ij}-\bar x_j)^2]} \Biggl \rangle 
    \label{eq2}
\end{equation}

where $\langle..\rangle$ represents the average over time, $x_{ij}$ indicates the x variable of the $i^{th}$ node in the $j^{th}$ layer and $\bar x_j$ is the mean value of $x_i$ in the $j^{th}$ layer .
The interlayer synchronization error is defined for nodes in different layers as

\begin{equation}
    S_{inter} =  \Biggl \langle \sqrt{\frac{1}{N}\sum_{i=1}^{N}[(x_{ij}- x_{ik})^2]} \Biggl \rangle 
    \label{eq3}
\end{equation}
where $x_{ik}$ indicate the x variable of the $i^{th}$ node in the $k^{th}$ layer where $k$ takes values $1,2$ or $3$ excluding $j$.

\begin{figure*}[!ht]
\includegraphics[width=1\textwidth]{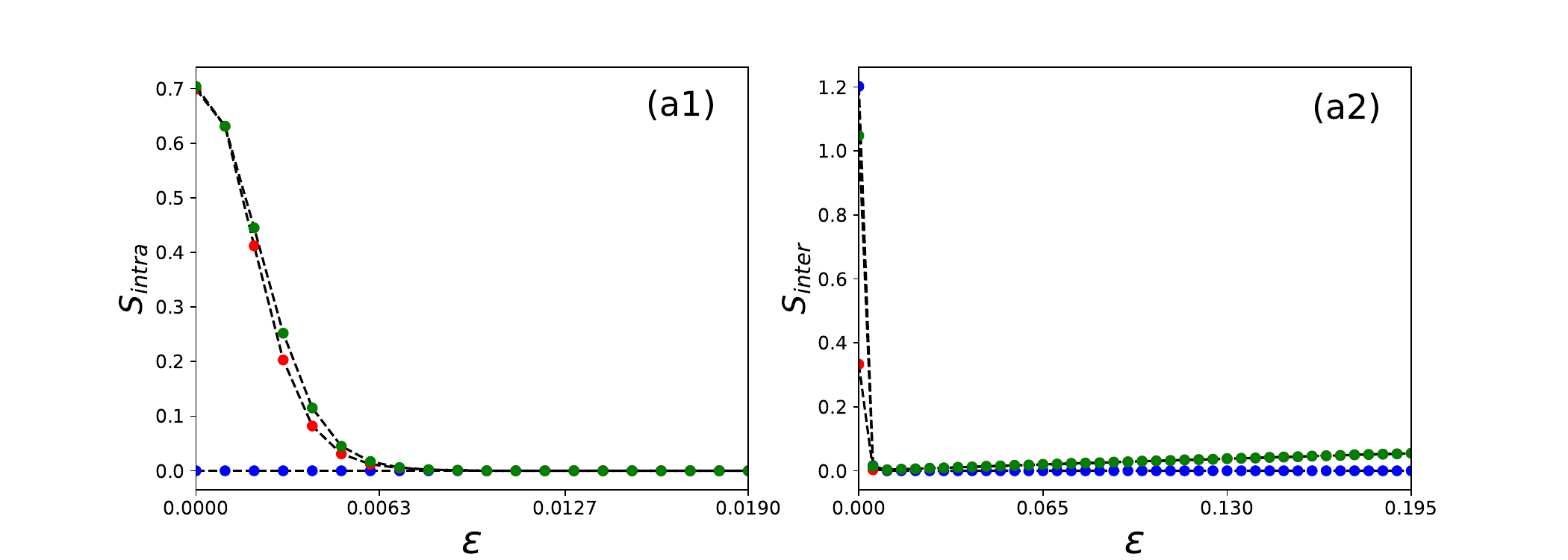}
\caption{Intra and inter-layer synchronization errors vs interlayer coupling strength $\epsilon$ for the three-layer multiplex network of 100 SL oscillators with unidirectional inter-layer coupling.    
(a1):For the chosen parameter values of $K1 = K3 = 0$, $K2 = 5$,  $P1 = P2 = P3 = 25$, $\omega = 2$, $\tau = 1$, the drive layer has $S_{intra} =0$, indicating it is in a state of intralayer synchronization(blue) while the response layers are not intralayer synchronized with $S_{intra}  \neq 0$ (red and green) for $\epsilon=0$. With the increase in the interlayer coupling ($\epsilon$), $S_{intra}$  goes to zero as the response layers achieve intralayer synchronization. (a2): For the parameter values of  $K1 = K2 = K3 = 5$,  $P1 = P2 = P3 = 25$, $\omega = 2$, $\tau = 1$, all the three layers have intralayer synchronization but no interlayer synchronization for $\epsilon=0$. With the increase of $\epsilon$, $S_{inter}$ between the response layers (blue) goes to zero, indicating the onset of relay synchronization. However, $S_{inter}$   between the drive and the response layers (red and green) first decreases and then increases with $\epsilon$. This is due to the amplification of oscillations on the response layers caused by the unidirectional feedback coupling from the drive layer.}
\label{Fig2}
\end{figure*}

For complete intralayer synchronization, $S_{intra} = 0$, while for complete inter layer synchronization, $S_{inter} = 0$. We plot the measure $S_{intra}$ for varying $\epsilon$ in Fig. \ref{Fig2} (a1). For the parameter values chosen, the drive layer L2 has complete intralayer oscillations (blue). It is clear that as $\epsilon$ is increased,  the intralayer synchronization error of both the response layers (red for L1 and green for L3) drops to zero, indicating the onset of synchronization in them.

To check for interlayer synchronization, we keep  $K1 = K3 = 5$ so that response layers have intralayer synchronization initially. Fig. \ref{Fig2} (a2) shows the variation in the inter-layer synchronization error as $\epsilon$ is increased. Here, the synchronization error between L1 and L2 is indicated in red, between L2 and L3 in green, and between L3 and L1 in blue.  It is clear that as $\epsilon$ is increased, both the response layers get synchronized to each other, achieving relay synchronization as shown in Fig. \ref{Fig2}(a2). Also, the synchronization error between the drive and response first decreases and then gradually increases with $\epsilon$ due to the amplification induced by the directional coupling.

\begin{figure*}[!ht]
\includegraphics[width=1\textwidth]{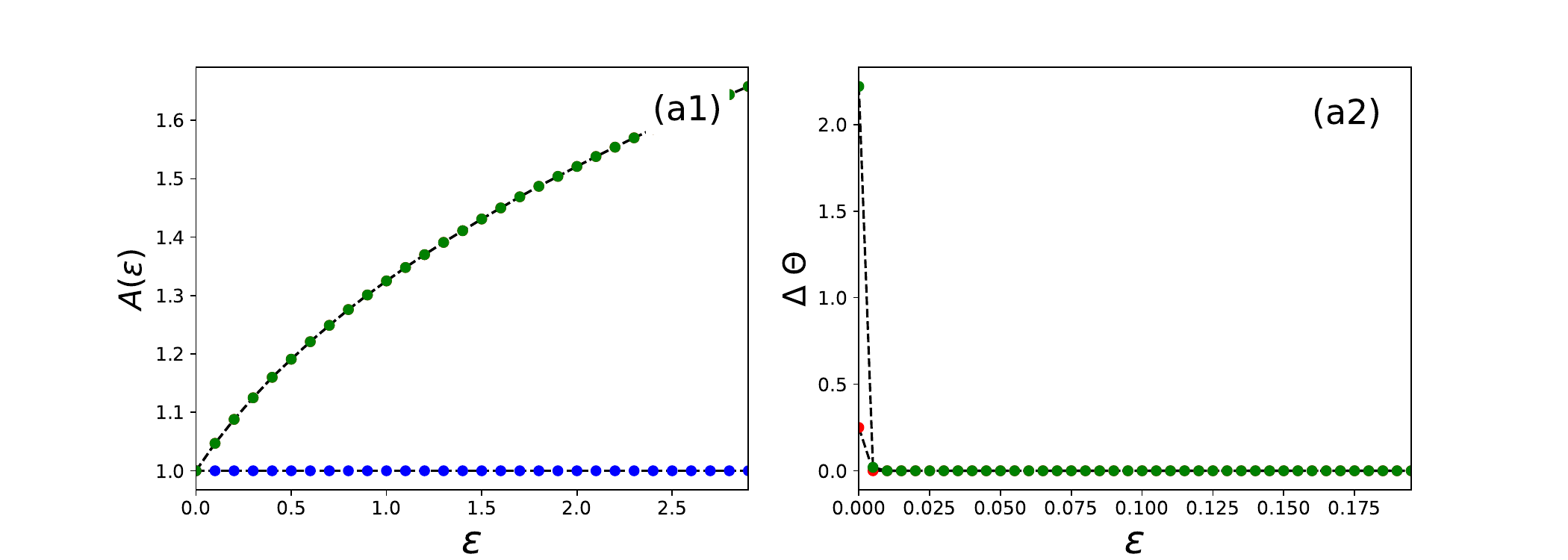}
\caption{(a1): Average amplitude measure, A($\epsilon$) vs strength of interlayer coupling $\epsilon$ for the three layers of the multiplex network of SL oscillators with unidirectional coupling.  We note the amplitudes of both the response layers L1 and L3 (green) are equal and increase with the value of $\epsilon$ while that of the drive L2 (in blue) remains the same. This indicates relay synchronization with amplification that can be adjusted by tuning $\epsilon$. (a2):  Average phase differences between the layers L1 and L2 (red) and that between L2 and L3 (green) as $\epsilon$ is increased. The phase differences decrease to zero, indicating phase synchronization between the drive and response layers even with amplification in the responses.
The other parameter values are kept as  $K1 = K2 = K3 = 5$, $P1 = P2 = P3 = 25$, $\omega = 2$, $\tau = 1$ and $N = 100$. }
\label{Fig3}
\end{figure*}

Now we study the role of directional coupling in amplifying the oscillations in the response layers. We define an average amplitude measure (relative to the drive) as A($\epsilon$) = $\frac{a(\epsilon)}{a(0)}$ where

\begin{equation}
    a(\epsilon)=\frac{1}{N}\sum_{i=1}^{N}[\langle x_{i,max} \rangle_t-\langle x_{i,min} \rangle_t]
    \label{eq4}
\end{equation}

and a($0$) is the measure for the drive layer. Fig. \ref{Fig3} (a1) shows the computed values of $A(\epsilon)$ in each layer with increasing $\epsilon$.  We note that as the strength of interlayer coupling ($\epsilon$) is increased, the amplitude of oscillations in both the response layers (green) increases, while that of the drive (blue) remains the same. We also compute the average phase difference, $\Delta \theta$, between oscillations in drive and response layers for varying values of $\epsilon$. As shown in Fig. \ref{Fig3} (a2), $\Delta \theta$ decreases to zero as $\epsilon$ is increased, indicating the onset of phase synchronization between the drive and response layers.

We repeat the above analysis for the case where the interlayer coupling is of diffusive type. For this in Eq. \ref{eq1}, we replace the feedback coupling term $\epsilon(x_{i2})$ by $\epsilon(x_{i2}-x_{ik})$ and $\epsilon(y_{i2})$ by $\epsilon(y_{i2}-y_{ik})$ for the x and y variables respectively, where $k = 1 \& 3$.  In this case, on adjusting $\epsilon$, both the response layers get synchronized with the drive resulting in complete synchronization in the entire multiplex network.

 \section{Effects of tuning dynamical time scales between layers}

\begin{figure*}
\includegraphics[width=0.9\textwidth]{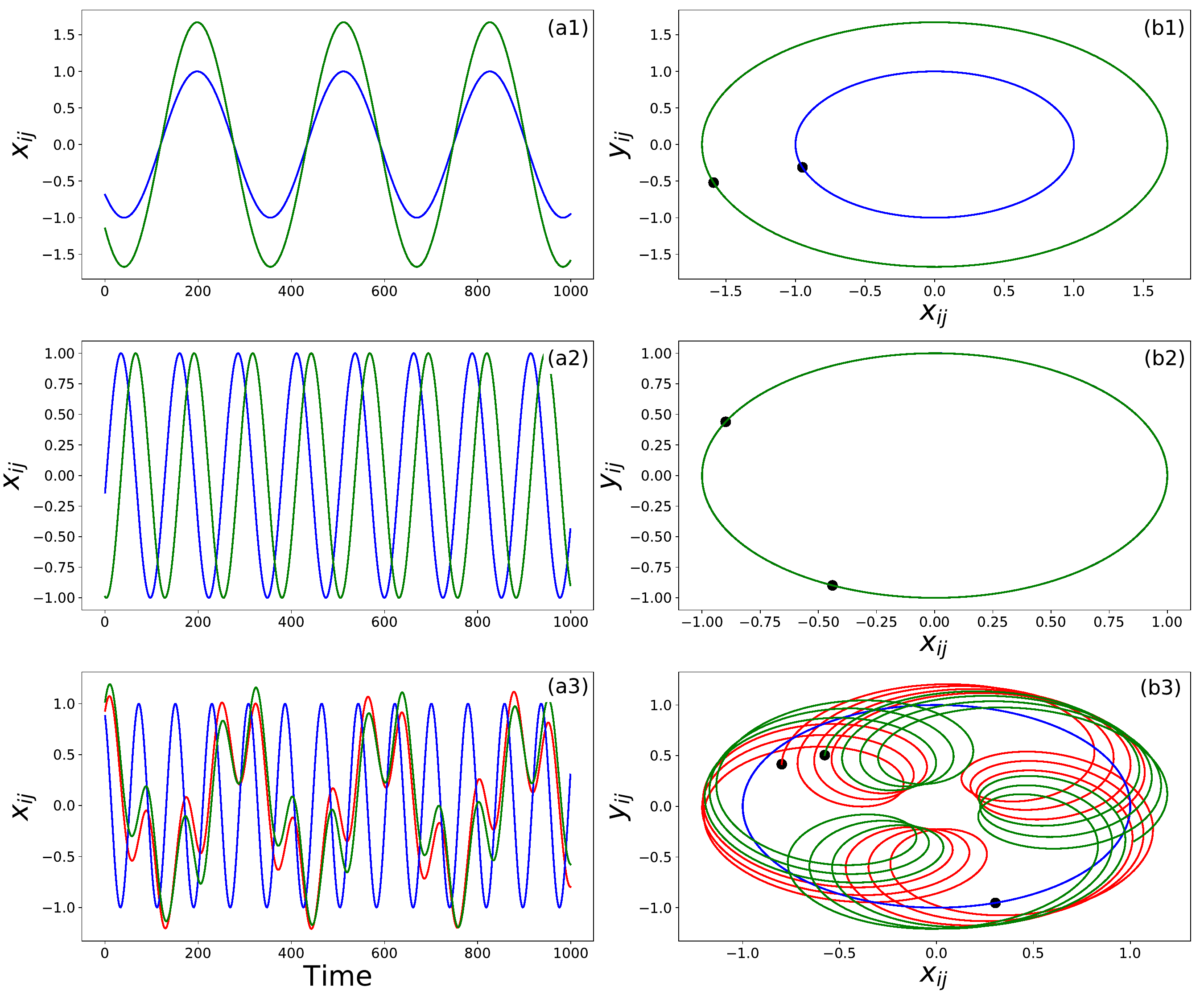}
\caption{Time series of the $x_{ij}$ variable with the index $i \in [1,N]$ and $j \in [1,2,3]$, and phase portraits in $x$ - $y$ plane for the three layer multiplex network of SL oscillators with unidirectional interlayer coupling of strength $\epsilon = 3$. The black dots on the trajectories indicate the values at the last time step. These show the different dynamical states induced in the response layers and the state of synchronization between the layers as the time scale mismatch between layers is varied. (a1) and (b1): Relay synchronization with amplification for the response layers (red and green) while phase synchronized with the drive layer (blue) for $\tau = 1$, (a2) and (b2): Responses layers are frequency synchronized with the drive with equal amplitude of oscillations and completely synchronized to each other as  $\tau$  is increased to  2.5, (a3) and (b3): Quasi periodic behaviour induced on the response layers for $\tau=4$ with frequency synchronization between them and functional relation with the drive layer. The other parameter values are: $K1=K2=K3=5$, $P1=P2=P3=25$, $\omega=2$ and $N=100$.}  
\label{Fig4}
\end{figure*}

In this section we present the effect of mismatch in the dynamical time scales between the drive and response layers. For this, we introduce a time scale difference between the layers by increasing the parameter $\tau$ ,  in Eq. \ref{eq1} and study the resulting changes in the dynamics of the response layers. We observe that the amplitude of oscillations in the response layers can be controlled by increasing $\tau$ and even made equal to that of the drive. With further increase in the value of $\tau$, the dynamics of the response layers change, and quasi-periodic oscillations set in, with only frequency synchronization between the response layers. We illustrate this scenario for chosen values of $\epsilon$ and $\tau$ in Fig. \ref{Fig4}.

For $\epsilon = 3$ and $\tau = 1$, in Fig. \ref{Fig4} (a1), we see the response layers are perfectly synchronized to each other, indicating relay synchronization and are phase synchronized with the drive layer.
The phase portrait of the three layers in Fig. \ref{Fig4} (b1) clearly indicates the amplification of oscillations for the response layers. As the drive layer is made to evolve on faster time scale, with $\tau = 2.5$, we observe that the amplitude of the oscillations in the response and drive layers are equal but are frequency synchronized with a constant phase difference between them as shown in Fig. \ref{Fig4} 
 (a2). This is clear from Fig. \ref{Fig4} (b2), where the phase portraits of the three layers overlap. With further increase of $\tau$, multiple frequencies are observed in the response layers, and the oscillations are quasi-periodic in nature, as shown in Fig. \ref{Fig4} (a3, b3) for $\tau=4$. Then, relay synchronization between the responses is lost and they are frequency synchronized to each other.

\begin{figure*}[!ht]
\includegraphics[width=0.7\textwidth]{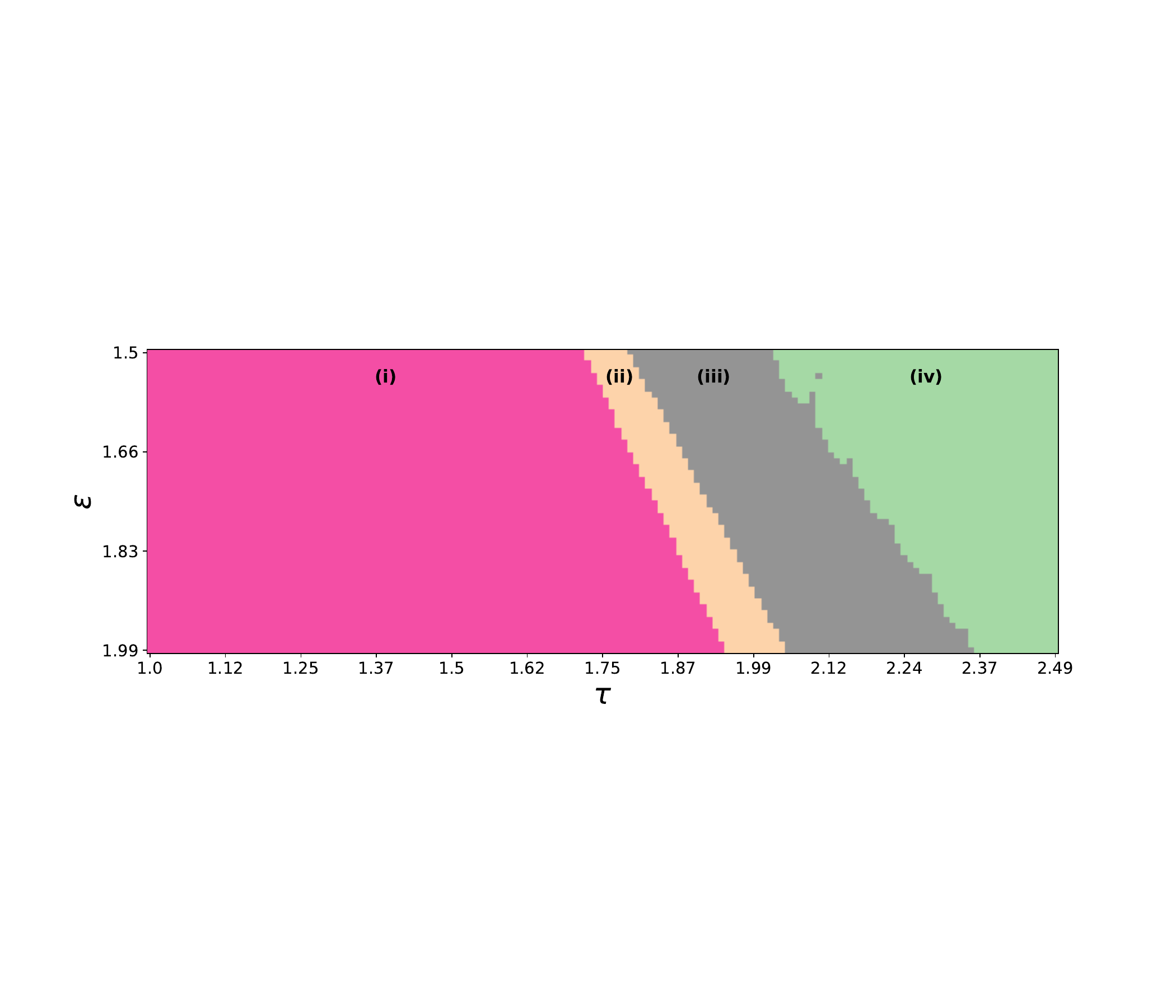}
\caption{Parameter plane of inter layer coupling strength $\epsilon$ vs  time scale mismatch $\tau$ for the three-layer multiplex network. The regions of various dynamical behaviors are indicated: (i): Amplification of oscillations on the response layers with respect to the drive,  (ii): Equal amplitude oscillations in both drive and response layers, (iii): Smaller amplitude oscillations in the response with respect to the drive, (iv): Quasi periodic behavior on the response layers. The other parameter values are as in Fig. \ref{Fig4}. 
}
\label{Fig5}
\end{figure*}

We indicate the possible dynamical states induced in the response layers in the parameter plane $\tau$ vs $\epsilon$ in Fig. \ref{Fig5}. We keep the other parameter values as $K1 = K2 = K3 = 5$, $P1 = P2 = P3 = 25$, $\omega = 2$ and $N = 100$ and study the nature of dynamics on varying $\tau$ and $\epsilon$ by computing average amplitude. The region marked as (i) indicates parameter values for which there is amplification in the response layer with respect to the drive. The average amplitudes of oscillations in the drive and the response layers are equal for the range of parameter values lying in the region (ii). In region (iii), the average amplitude of the response is less than that of the drive, and in region (iv), quasi-periodic states are observed in the response layers.

\begin{figure}[!ht]
\includegraphics[width=0.48\textwidth]{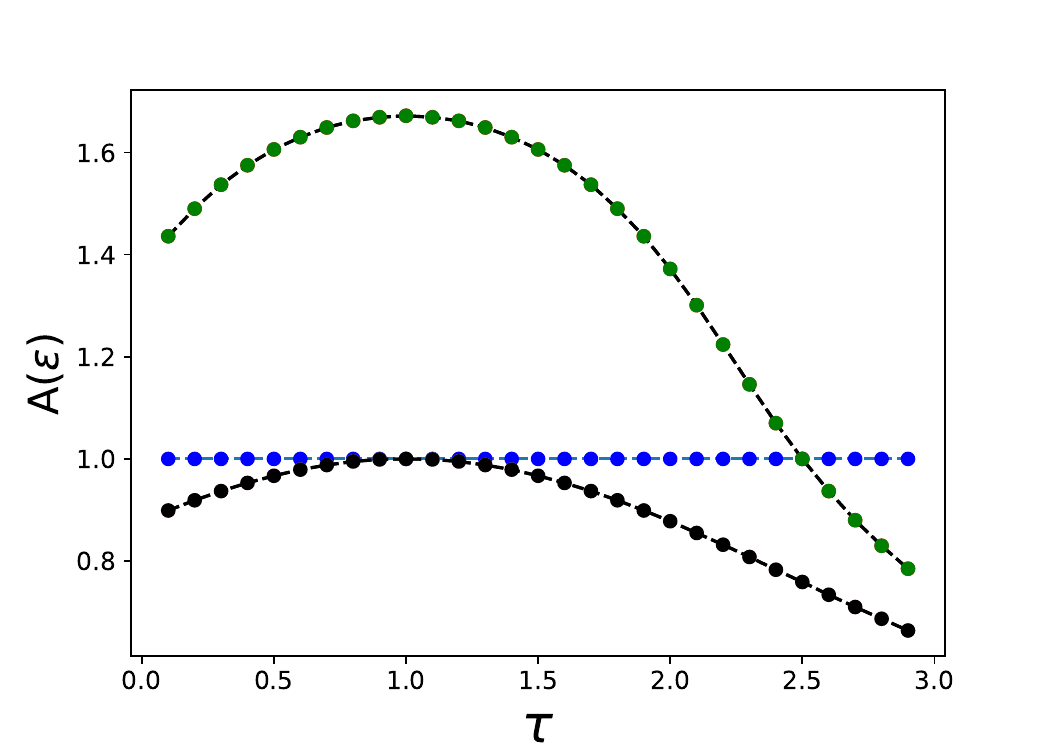}
\caption{Variation in the average amplitude measure, A($\epsilon$), with the time scale mismatch $\tau$ fixing $\epsilon=3$ in the three-layer multiplex network for two types of interlayer coupling. Here, the blue curve shows the average amplitude of the drive layer, and the green( black) is that of the synchronized response layers with unidirectional coupling of feedback (diffusive) type. The other parameter values are: $K1 = K2 = K3 = 5$, $P1 = P2 = P3 = 25$, $\omega = 2$, and $N = 100$. For feedback coupling, amplification can be achieved, while for the diffusive type of interlayer coupling, the amplitude is less than that of the drive, except for $\tau=1$. In both cases, the amplitude of oscillations in response layers can be controlled by tuning $\tau$.}
\label{Fig6}
\end{figure}

The variation in the relative average amplitude, A($\epsilon$), for L1 and L3 with $\tau$ is shown in Fig. \ref{Fig6} for both feedback coupling and diffusive coupling between the layers. In the case of directional feedback coupling for $\epsilon = 3$, when both the drive and the response layers have the same time scales, the average amplitude of the response is larger than the drive. As a time scale difference is introduced between the response and the drive (both $\tau > 1$ and $\tau < 1$), the average amplitude of the responses decreases, becoming equal at $\tau = 2.5$. In the case of diffusive interlayer coupling, as the drive layer is made faster ($\tau > 1$) or slower ($\tau< 1$) the response layers show relay synchronization along with frequency synchronization and decreased average amplitude when compared to the drive as depicted in Fig. \ref{Fig6} (black).  So, introducing a dynamical time scale difference between the drive and response, in this case, can make the system go from complete synchronization to relay synchronization but with a lower amplitude of oscillations in response layers.

In addition to the time scale mismatch between layers, we also consider the case with a mismatch in the parameters of the intrinsic dynamics, $\omega$, of SL oscillators between layers. Thus, with $\omega_{drive} = 2$, $\omega_{response} = 1.5$ and other parameters kept as $K1 = K2 =
K3 = 5$, $P1 = P2 = P3 = 25$, $\tau = 1$, $N = 100$ and $\epsilon = 3$, for feedback type of interlayer coupling, we observe relay synchronization with amplification of the response layers, but they have only frequency synchronization with the drive, unlike the previous case where phase synchronization is observed. 
As $\tau$ is varied for a fixed $\epsilon = 3$, similar to the above observations, tuning of the amplitude of oscillations of the response relative to that of the drive is possible. This is then followed by quasi-periodic behaviour for higher $\tau$ values. In this case, if the interlayer coupling is of diffusive type, we observe relay synchronization of the responses instead of complete synchronization due to the parameter mismatch along with frequency synchronization with the drive, and the amplitude of oscillations of the response layers can be controlled by tuning $\tau$.

\section{Conclusion}

While relay synchronization is one of the active research areas in the context of complex networks, most of the reported studies till now consider bidirectional coupling in the multilayer framework. In this study, we explore the effects of unidirectional interlayer coupling in a three-layer multiplex network of Stuart Landau oscillators, with the middle layer L2 being the drive and the remote layers L1 and L3 being the responses. We analyze how different coupling strategies lead to the development of relay synchronization on the response networks.

The interactions between layers in the multiplex networks enable the system to achieve different dynamical states. The specific type of interlayer coupling (feedback or diffusive) decides the synchronization between layers, while the presence of time scale or parameter mismatches significantly affects the collective behaviour.  This allows for the control of dynamics in response layers by tuning the parameters in the drive layer.

When all the layers evolve at the same dynamical time scale and with identical systems, the unidirectional feedback from the drive causes amplification of oscillations in the response layers along with relay synchronization that is in phase with the drive. We show that the amplification can be controlled by adjusting the strength of interlayer coupling.

By introducing appropriate time scale differences between the layers using the mismatch parameter $\tau$, we can control the amplitude of oscillations in the response layers, and they can be made equal to that of the drive with a phase difference. With further increase in $\tau$, the amplitude of oscillations of the response becomes smaller than the drive, finally leading to quasiperiodic behaviour in the response layers. However, as quasi-periodic behaviour emerges, relay synchronization is lost, with only frequency synchronization between the response layers.

In the case of unidirectional diffusive coupling between identical layers, we obtain synchronization between the drive and responses, leading to complete synchronization of the entire network. When a time scale mismatch is introduced, the amplitude of oscillations of the response layers decreases and results in frequency synchronization with the drive layer

We also study the dynamics on the response networks when there is a parameter mismatch in the response compared to the drive layer. In this case, for both feedback and diffusive interlayer coupling, the response layers get completely synchronized with each other.  In this context, we note that a few studies have been reported on the generalized synchronization of complex networks \cite{Shang2009, NING2019104947, ning2015driving}. The framework of multiplex network used in the study can be used as the auxiliary system approach to explore generalized synchronization between different layers of
complex networks where one network drives two similar networks that get synchronized.

The study can be extended further to multiple response layers with different dynamics and topologies that can induce various levels of synchronous dynamics and spatiotemporal patterns. Some promising directions for future research include networks consisting of diverse dynamical units, different coupling mechanisms including higher order coupling,  delay in couplings, varied network topologies, and also introducing adaptive mechanisms within the network that could provide insights into how networks self-regulate to maintain synchronization.  Such extensions of the present study may require more sophisticated modelling and analytical techniques. However, validating the results of these studies with empirical data from real-world systems like neuronal networks and smart grids will be value-adding but challenging.

The present study provides an understanding of the occurrence and control of collective behaviour in multiplex networks with unidirectional interlayer interactions. It can lead to possible mechanisms for remotely controlling the dynamics to desired states on complex networks by adjusting the strength of driving and the parameters of the driving network. This will have applications in controlling layers of networks in smart grids as well as in modelling remote synchronization in neuronal networks.

\section{Data Availability and software Packages}

All the data used in the study are generated by integrating the system of equations for the dynamics of the multiplex network.  The computations are done using Python codes adapted to the context with publicly available packages such as matplotlib.pyplot , scipy.integrate, networkx.

\bibliography{reference}

\end{document}